\documentclass[letterpaper,twocolumn,10pt]{article}
\usepackage{usenix,url,amsfonts,amsmath,amsthm,multirow,amssymb,bmpsize}
\usepackage{graphicx}
\usepackage{subcaption}
\usepackage{enumitem}
\usepackage{booktabs}
\usepackage{listings}  
\usepackage{url}

\usepackage{breakurl}
\usepackage[breaklinks]{hyperref}
\usepackage{xcolor}
\setlist{noitemsep}
\graphicspath{{figs/}}
\DeclareGraphicsExtensions{.png}

\usepackage{adjustbox}
\usepackage{array}
\usepackage{float}
\usepackage[margin={1.5cm,1.5cm}]{geometry}

\newcommand\mytt[1]{\texttt{\small{#1}}}

\newcommand\bktcolumns{\texttt{\small{bkt:columns~}}}
\newcommand\bktranges{\texttt{\small{bkt:ranges~}}}
\newcommand\pidcounts{\texttt{\small{bkt:pid\_cnts~}}}
\newcommand\noisycnt{\texttt{\small{bkt:noisy\_cnt~}}}
\newcommand\suppress{\texttt{\small{bkt:suppress~}}}
\newcommand\lowval{\texttt{\small{low\_val}}}
\newcommand\highval{\texttt{\small{high\_val}}}
\newcommand\onedim{\texttt{\small{1dim~}}}
\newcommand\ndim{\texttt{\small{ndim~}}}
\newcommand\nonedim{\texttt{\small{n-1dim~}}}

\newif\ifblind
\blindfalse

\ifblind
\newcommand\sd{SynDiffix~}
\newcommand\sdx{SynDiffix}
\newcommand\sdurl{github.com/a/blinded/url}
\newcommand\sdpyurl{github.com/a/blinded/url2}
\newcommand\datasetsurl{blinded.com/a/blinded/url}
\else
\newcommand\sd{SynDiffix~}
\newcommand\sdx{SynDiffix}
\newcommand\sdurl{github.com/diffix/syndiffix-fs}
\newcommand\sdpyurl{github.com/diffix/syndiffix}
\newcommand\datasetsurl{gitlab.mpi-sws.org/francis/test-datasets}
\fi

\newcolumntype{R}[2]{%
    >{\adjustbox{angle=#1,lap=\width-(#2)}\bgroup}%
    l%
    <{\egroup}%
}

\begin{document}

\date{}

\title{\sdx: More accurate synthetic structured data}

\author{
Paul Francis$^{\dag}$ \quad Cristian Berneanu$^{\S}$ \quad Edon Gashi$^{\S}$ \\
$^{\dag}$Max Planck Institute for Software Systems (MPI-SWS), Germany\\
$^{\S}$Open Diffix\\
francis@mpi-sws.org, \{cristian, edon\}@open-diffix.com
}

\maketitle

\begin{abstract}
  This paper introduces \sdx, a mechanism for generating statistically accurate, anonymous synthetic data for structured data. Recent open source and commercial systems use Generative Adversarial Networks or Transformed Auto Encoders to synthesize data, and achieve anonymity through overfitting-avoidance. By contrast, \sd exploits traditional mechanisms of aggregation, noise addition, and suppression among others. Compared to CTGAN, ML models generated from \sd are twice as accurate, marginal and column pairs data quality is one to two orders of magnitude more accurate, and execution time is two orders of magnitude faster. Compared to the best commercial product we measured (MostlyAI), ML model accuracy is comparable, marginal and pairs accuracy is 5 to 10 times better, and execution time is an order of magnitude faster. Similar to the other approaches, \sd anonymization is very strong. This paper describes \sd and compares its performance with other popular open source and commercial systems.
\end{abstract}

\section{Introduction}

Anonymization of \textit{structured data} is an important and active research problem. The problem is to release structured data to untrusted parties in such a way that the parties can derive statistical information from the data without being able to learn about specific individuals in the data.

Since the publication of CTGAN in 2019~\cite{xu2019modeling} there has been increased interest in generating anonymous~\textit{synthetic data}. While some of this interest is from the academic community (see~\cite{figueira2022survey} for a good survey), a substantial amount is commercial. \cite{synStartups} lists 45 startups with structured synthetic data products, most of them using Generative Adversarial Networks (GAN) or Transformed Auto-Encoders (TAE).

A huge practical advantage of synthetic data is that it ``looks like'' the original data. A row in the synthetic data behaves like a row in the original data (though there is not a one to one mapping). The synthetic data can be plugged into any analytic software tools just as the original data can. 

This is not the case with systems that generate aggregate data such as counts, sums, averages, and so on. To compensate for this lack of flexibility, some aggregate anonymizing systems open a \textit{dynamic} query interface which gives users some control over processing and selecting data~\cite{pinq,DBLP:journals/corr/abs-1809-07750, francis2022diffix, tableBuilder}. This is in contrast to \textit{static} aggregate approaches, where the data owner decides in advance what aggregates will be released.

A major difficulty with dynamic query systems is that they open up a large attack surface. The query operates on the original data itself, and the result is aggregated and anonymized. The differential privacy query systems PINQ~\cite{pinq} and Chorus~\cite{DBLP:journals/corr/abs-1809-07750} were both open to query side channel attacks that effectively allowed attackers to recreate the original data~\cite{dp-under-fire,boenisch2021side}. Early versions of Diffix~\cite{diffix-cedar-tr}, which offered a rich subset of SQL, were found to have vulnerabilities~\cite{cohen2019linear,montjoyeAttack}, as was an early version of the Australian Bureau of Statistics system DataBuilder~\cite{asghar2019averaging}.

Synthetic data systems have a much smaller attack surface because any ``queries'' the user makes are run on the anonymized synthetic data, not on the original data.

Of course synthetic data is not a panacea. Simply because any query or analysis can be run over synthetic data does not mean that it produces accurate results.

This paper presents \sdx, an anonymizing structured data synthesizer. Compared to CTGAN, as implemented by SDV~\cite{sdvCtgan}, the median performance of \sd has over two times better ML efficacy, \textbf{12x more accurate} single-column (marginals) quality, \textbf{80x more accurate} column pairs quality, and executes \textbf{two orders of magnitude faster}. Compared to the best commercial product we tested (MostlyAI), \sd has comparable ML efficacy, \textbf{5x and 13x more accurate} marginals and pairs quality respectively, and executes \textbf{one order of magnitude faster}.

Like ML-based approaches, \sd is strongly anonymous, with equal or better privacy properties.

After discussing related work in Section~\ref{sec:related}, we present the design of \sd and describe why the design is anonymous (Section~\ref{sec:design}). In Section~\ref{sec:setup} we describe the experimental setup, followed by a comparison of \sd with other open source and commercial mechanisms for data quality, ML efficacy, and execution time (Sections~\ref{sec:quality}, \ref{sec:ml-efficacy}, and \ref{sec:elapsed} respectively). In Section~\ref{sec:privacy} we discuss anonymity measures and present measures for \sd the other approaches. Finally Section~\ref{sec:future} summarizes and presents future work.

There are two open-source implementations of \sdx. One is built from F\#\footnote{\sdurl} and is the implementations used for the experiments in this paper. The other is pure python\footnote{\sdpyurl}. Functionally the two are equivalent, but the pure python version is around five times slower.

The contribution of this paper is a new approach to synthetic structured data that has equal or better ML efficacy, many times better data accuracy, and executes many times faster. It has strong anonymity, comparable to ML-based systems, and better than Synthpop.

\section{Related work}
\label{sec:related}

The literature on anonymization of structured data for statistical disclosure is vast, constituting thousands of papers and spanning five decades.

Traditional \textit{statistical disclosure} techniques aim to release statistically accurate data while preserving anonymity. Mechanisms include removal of personally identifying information (PII), aggregation, sampling, suppression (of outliers or aggregates with low counts), and adding noise (to aggregates or to individual values). A good survey article for these kinds of mechanisms is~\cite{matthews2011}. \sd uses a number of these techniques in combination, borrowing heavily from Diffix~\cite{francis2022diffix}

A second broad area of research concerns generating \textit{synthetic} data. The primary goal here is to generate data that has the ``look and feel'' of the original data, and can take the place of the original data in applications that use the data. Use cases include data for testing purposes, augmenting the original data in cases where there is not enough data or where certain properties need to be changed or emphasized (fairness, or rare events like fraud), or replacing the original data. Goals include statistical accuracy, machine learning (ML) efficacy, or capturing other properties of the data. Anonymization per se is often not a primary goal, but nevertheless is a side effect of data synthesis. Good survey papers include~\cite{lu2023machine} and~\cite{figueira2022survey}.

There are two broad approaches to synthetic data. One is to model the data and then sample from the model. Commonly used models are decision trees~\cite{reiter2005using} and Bayesian networks~\cite{zhang2017privbayes}. A more recent broad approach is to use generative networks, particularly transformed auto encoders (TAE) and generative adversarial networks (GAN)~\cite{xu2019modeling, rajabi2022tabfairgan}. \sd takes a model-and-sample approach using decision trees customized to preserve anonymity.

Something of a breakthrough was achieved in 2019 with the development of CTGAN (Conditional Tabular GAN)~\cite{xu2019modeling}. There has been significant development based on CTGAN, but most of it commercial rather than academic: there are literally more startups based on CTGAN than follow-up papers (see~\cite{synStartups}). While it is obviously problematic to include confidential proprietary technologies in an academic paper, the fact remains that the startup implementations we tested (Gretel and MostlyAI) both perform substantially better than CTGAN and so should not be ignored.

The core element of \sd is a bucket with sticky noise, snapped ranges, and suppression. This is taken directly from Diffix~\cite{francis2022diffix}. Whereas Diffix requires the user to specify the ranges in SQL, \sd discovers them automatically. Early versions of Diffix~\cite{diffixBirch,diffix-cedar-tr} offered a rich set of SQL features, leading to a large attack surface and subsequent vulnerabilities~\cite{cohen2019linear,montjoyeAttack}. Anonymization systems with a rich query interface have a checkered history. The differential privacy systems PINQ~\cite{pinq}, Airavat~\cite{roy2010airavat}, and Chorus~\cite{DBLP:journals/corr/abs-1809-07750} all have severe side-channel vulnerabilities~\cite{dp-under-fire, boenisch2021side}.  Note that private clean rooms of Google's Ads Data Hub~\cite{ads-data-hub} and the Amazon Marketing Cloud~\cite{amazon-marketing-cloud} also have rich SQL interfaces. We are not aware of any reports of vulnerabilities in these systems.

By contrast, \sd offers a simple table-in-table-out interface, and therefore a much smaller attack surface. To the extent that query processing is required, it can be done on the synthetic data itself (so long as the synthesis captures the desired properties).

The closest prior work to \sd that we are aware of is Synthpop~\cite{nowok2016synthpop}. Synthpop is a collection of synthetic data tools implemented in R. A key tool is is based on CART~\cite{reiter2005using}, which like \sd uses decision trees to model the data. Synthpop has a couple of optional features geared towards strengthening anonymity, such as smoothing sampled continuous data and requiring a minimum number of rows in the leaves of trees (\mytt{cart.minbucket}). With its use of Diffix-style buckets, \sd has more and stronger anonymization mechanisms.

%
%
%
%
%
%
%
%
%
%
%


\section{\sd Design}
\label{sec:design}

Because of space constraints, this description focuses on the key concepts of \sdx. Missing details can be found in the Github repository.

\sd implements specialized decision trees where each node in the tree has basic strong anonymity properties. Synthetic data is generated by random sampling from the nodes.

\subsection{\sd interface}

The interface to \sd is:

\vspace{-2ex}
\begin{figure}[H]
\begin{small}
\begin{tabular}{rll}
   \textbf{Input:} &   \textit{original table} & \mytt{\# columns by rows} \\
          &   \textit{protected entities} & \mytt{\# NULL or columns} \\
          &   \textit{ML target column} & \mytt{\# NULL or column} \\
          &   \textit{columns to synthesize} & \mytt{\# defaults to 'all'} \\
          &   \textit{other settings} &  \\
   \textbf{Output:} &   \textit{synthetic table} & \\
\end{tabular}
\end{small}
\end{figure}
\vspace{-3ex}

The current implementation supports text, numeric, datetime, and boolean columns in the \textit{original table}. \sd is not limited to these types.

The \textit{protected entity} (PE) is the thing being protected. Normally this is an individual person, but can be a person's device or anything else. \sd can protect more than one type of entity. Examples include sender/receiver, patient/doctor, and IP address/email address. \sd requires that there be a column containing unique identifiers (PID) for the PE. The exception to this is when there is a single type of PE, and each PID has only one row (e.g. survey data). In this case, PE can be \mytt{NULL}, and \sd will internally generate a PE column with one distinct PID per row. Otherwise, the PE columns are listed in the API.

If the use case is to predict a given attribute with an ML model, then specifying the \textit{ML target column} will improve those models.

\sd ignores any columns not listed in \textit{columns to synthesize}, and does not include them in the output. This API parameter is superfluous in that one could simply delete the columns from the table in advance. We put it in the API, however, to emphasize a key difference between the usage style of SynDiffix versus other synthesizers (\S\ref{sec:usage-difference}).

The output is the synthesized data. It is important to note that the PID columns in the output do not contain useful information. As a result, \sd does not accurately synthesize event (time series) information, for instance relationships between events in time, or the distribution of rows among PIDs. Accurately synthesizing events is certainly possible, but is left for future work.

\subsubsection{Usage difference versus other synthesizers}
\label{sec:usage-difference}

\begin{figure}
\begin{center}
\includegraphics[width=0.95\linewidth]{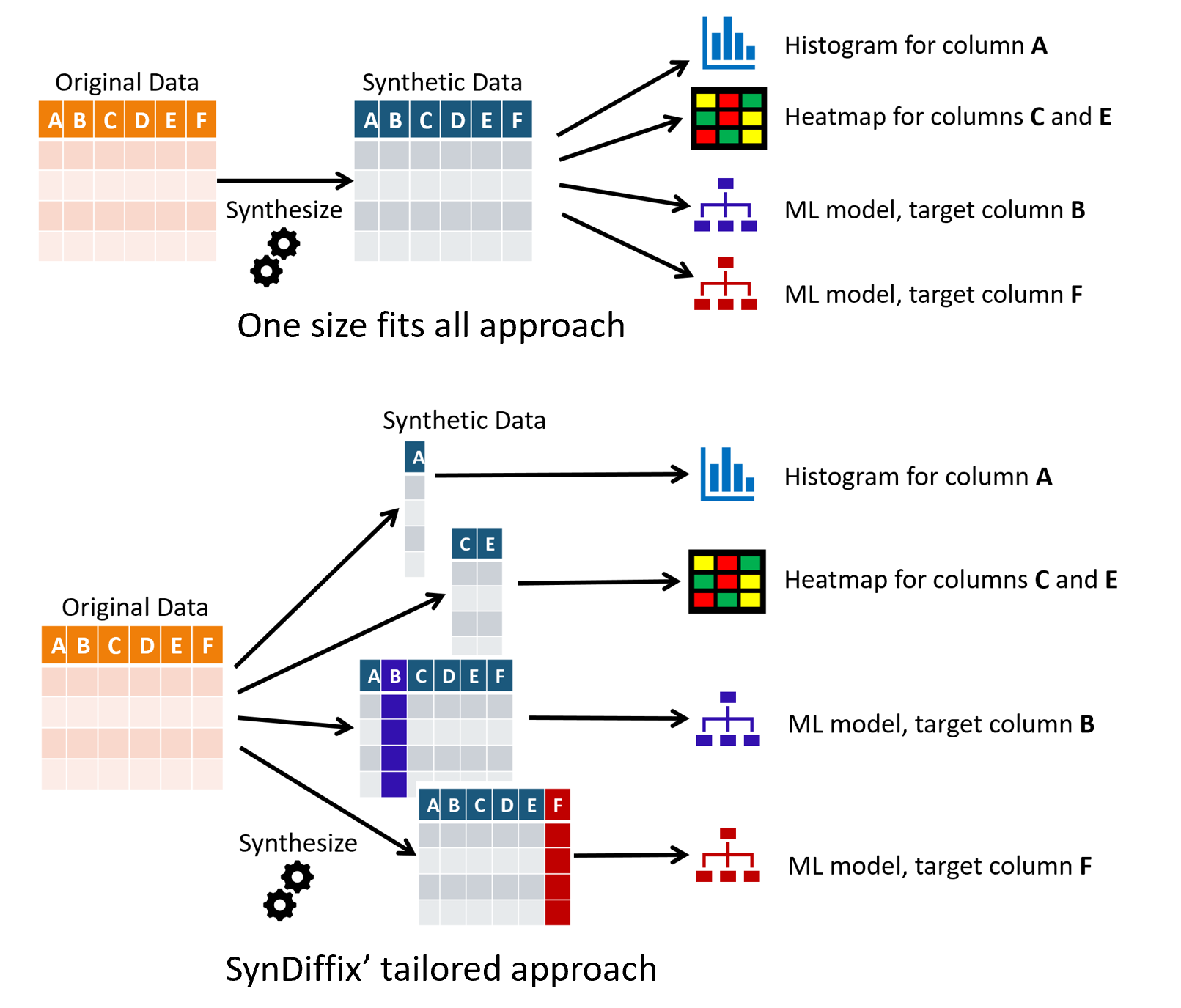}
\caption{One size fits all versus \sdx' tailored usage approach. \textnormal{Although one size fits all is more convenient, the tailored approach leads to more accurate results for each given use case.}}
\label{fig:usage}
\end{center}
\end{figure}

The implied usage style of existing synthesizers is \textit{one-size-fits-all}: the complete table is synthesized, and any downstream analytic task operates over the synthesized table (see Table~\ref{fig:usage}). The usage style of \sd is to tailor the synthesis to the downstream task.

For example, a fundamental property of any anonymization system is that the more attributes one tries to capture, the worse the resulting data quality for any given attribute. If one is interested for example in the correlation between columns A and B, and separately between columns A and C, then the best results are achieved by synthesizing only columns A and B, and separately synthesizing only columns A and C.

\sd fully embraces the idea of generating tailored syntheses for different downstream analytic tasks. The API lets users easily select only the columns of interest. The API also allows users to specify the target column where the downstream task is ML prediction of that column. This leads to better results than disregarding the target column during synthesis. As we add additional attributes of interest, for instance event sequences in time series data, the ability to tailor synthesis for those attributes will maximize quality.

Of course with existing synthesizers it is possible to select different columns for different syntheses simply by removing columns before running the synthesis. It is important to note, however, that repeated syntheses of the same column tends to erode privacy. The underlying randomness of the synthesis process tends to expose new information with each execution.

\sd doesn't have this property. Its use of snapping and sticky noise (see \S~\ref{sec:core-concepts}) limits the exposure of any given piece of data regardless of how many syntheses it appears in.

We are not aware of any studies that examine privacy loss of existing synthesizers due to repeated exposures. Intuitively it seems unlikely to lead to serious risks in practical settings, but we don't know.

\subsection{The anonymized bucket is the core building block}
\label{sec:core-concepts}

\sd synthesizes data by building a set of regression trees of one or more dimensions (columns). A node in a tree is called a \textit{Diffix bucket} (or just \textit{bucket} for short). A Diffix bucket is an aggregate of multiple rows, and has strong anonymizing properties. Diffix buckets are the basic building block of \sdx. Row-level \textit{microdata} is synthesized from these buckets. The elements of a Diffix bucket are listed in Table~\ref{tab:bucket}.

\begin{table}
\begin{center}
\begin{small}
\begin{tabular}{ll}
   \toprule
   \multicolumn{2}{l}{\textbf{Basic parameters}} \\
   \bktcolumns & Bucket columns \\
   \bktranges & Column ranges \\
          & Each range: [\lowval, \highval), \\
          & or single value \\
   \pidcounts & The row count for each distinct PID \\
   \midrule
   \multicolumn{2}{l}{\textbf{Derived parameters}} \\
   \noisycnt & The noisy count of rows \\
   \suppress & Whether or not to suppress bucket \\
   \bottomrule
\end{tabular}
\end{small}
\end{center}
  \caption{Parameters of an anonymous bucket. \textnormal{}}
  \label{tab:bucket}
\end{table}

A bucket may have one or more dimensions, each associated with a column. Each dimension has a \textit{range}, which may be a [\lowval, \highval) pair, or a single value. In the implementation, we cast non-numeric values as numeric values for use as ranges (\S~\ref{sec:casting}).

The \pidcounts gives the number of rows associated with each distinct PID for each type of PE.

These basic parameters are generated from the original table. From the basic parameters, the \noisycnt and \suppress flag are derived. The \noisycnt is the perturbed count of rows assigned to the bucket, and determines how many synthetic rows to build. \suppress is \mytt{TRUE} when there are too few distinct PIDs associated with the bucket.

Two key attributes of buckets are \textit{sticky noise} and \textit{snapping}.

\textbf{Sticky noise:} Each bucket is subject to a number of pseudo-random decisions, for instance how much noise to add to counts and whether or not to suppress the bucket. Buckets are \textit{sticky} in the general sense that the ``same bucket'' will produce the ``same pseudo-random decision''. The purpose of sticky noise is to prevent the randomness from being averaged away due to repetitions of the same bucket. Sticky noise is implemented by seeding a pseudo-random number generator with elements taken from the bucket itself: the table name, column names, and ranges.

\begin{figure}
\begin{center}
\includegraphics[width=0.85\linewidth]{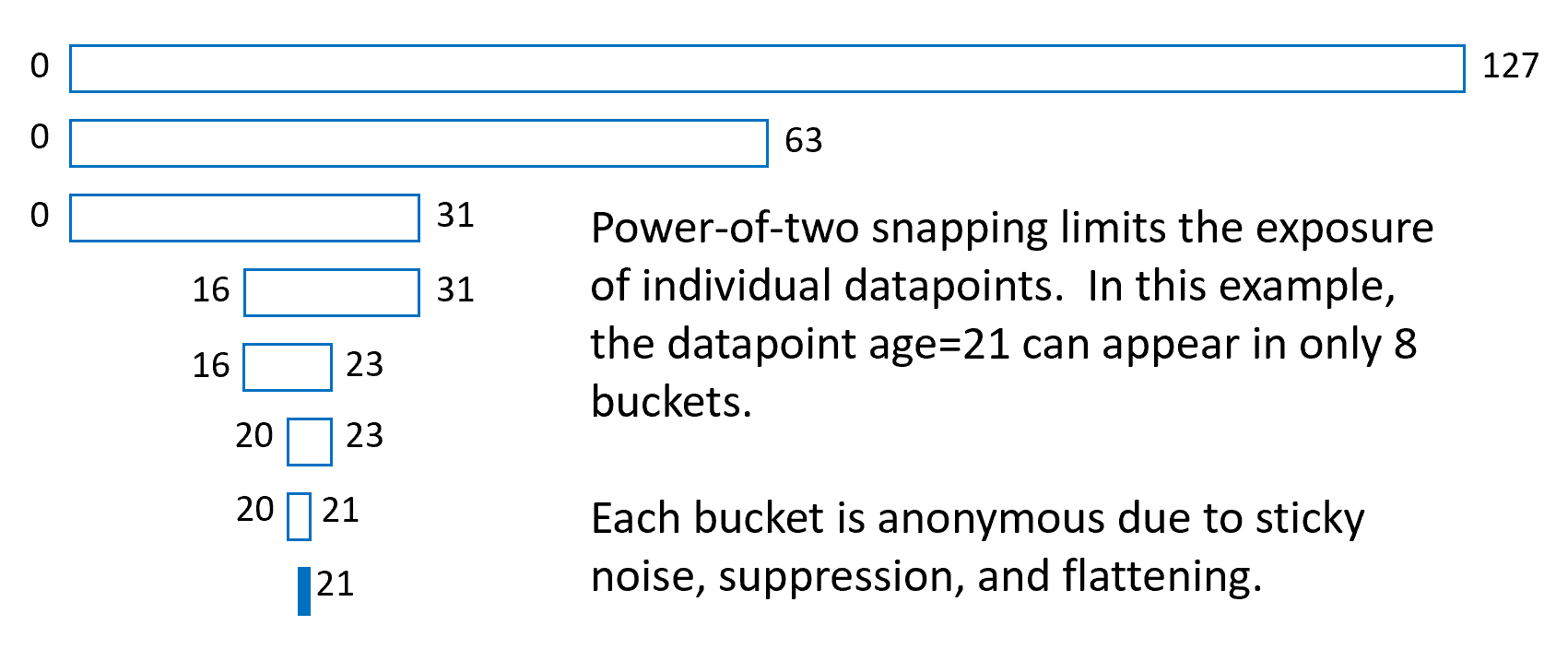}
\caption{Power-of-2 snapping. \textnormal{}}
\label{fig:snapping}
\end{center}
\end{figure}

\textbf{Snapping:} Bucket ranges are limited to a pre-determined set of sizes and offsets, namely power-of-two sizes and offsets. This limits any given datapoint to a small number of buckets. For instance, Figure~\ref{fig:snapping} shows that the datapoint \mytt{age=21} can appear in at most eight different one-dimensional buckets.

Taken together, sticky noise and snapping limit the exposure of any given data point. There are a limited set of buckets, and because each bucket is sticky, multiple exposures in the same bucket due to multiple syntheses does not eliminate the noise associated with the bucket.

Buckets have four additional anonymizing properties, \textit{proportional noise}, \textit{flattening}, \textit{suppression}, and \textit{PID noise}. These mechanisms are implemented exactly as with Diffix Elm buckets, the details of which can be found in \S 3.5 of \cite{francis2022diffix}.

\textbf{Proportional noise:} The amount of noise added to the true bucket count is proportional to the number of rows contributed by individual PEs. The amount of noise is proportional to the average row contribution of the top few contributors, after flattening. The exact number of PEs from which this average is taken is itself a sticky noisy value, typically ranging from 3 to 5 of the top contributing PEs.

\textbf{Flattening:} If a given PE contributes an extreme number of rows to a bucket, relative to other PEs, then without flattening the presence or absence of the extreme contributor could be detected from the amount of noise itself. Flattening reduces the row count associated with the top one or two \textit{extreme contributors} to match the average row count of the next few \textit{top contributors}. The resulting proportional noise is therefore effectively derived from multiple PEs rather than a single PE. (Again, the exact number of extreme contributors and top contributors used in the calculation are themselves sticky noisy values.)

\textbf{Suppression:} To prevent the release of data values associated with a single PE, buckets that have too few distinct PIDs are suppressed in the sense that they are not used to produce microdata. For example, suppose that there is only one individual with \mytt{age>=100} in a table, and that that person has \mytt{age=103}. The buckets \mytt{[103]}, \mytt{[102,103]}, and \mytt{[100,103]} may all be suppressed due to having too few contributing PEs, thus hiding the presence of that individual. Note that the rows from suppressed buckets are not removed from the synthetic data, but rather are merged with parent buckets. In other words, suppression leads to loss of precision, not loss of data.

\textbf{PID Noise:} Each bucket has a second proportional noise value. This value is seeded by a hash composed of all distinct PIDs. The primary purpose of this additional noise is to mitigate what can be learned from changes in dynamic data. With sticky noise alone, two repeated queries will have the same noise value, and so any difference in the query results are due to changes in the underlying data. PID noise mitigates this problem by masking the exact amount of change in cases where there is some change (if no change, then the result of the two queries remains the same).

One possible additional mechanism to further mitigate the problem of dynamic data change is to divide time into epochs and exclude any changes since the end of the last epoch. Another would be to add yet another noise value seeded by the epoch parameters themselves in order to mask the fact that no change took place between epochs. This is future work.

\subsection{Anonymous trees from anonymous buckets}

The core anonymity properties of \sd derive from Diffix buckets. A key idea is that, so long as the anonymity properties of buckets are preserved, any structures built out of buckets, and any microdata derived from buckets, are also anonymous.

There are many different designs that can be built on this key idea. \sd as described in this paper is one such design. It is almost certainly not the best, and we suggest improvements here and there.

\begin{figure}
\begin{center}
\includegraphics[width=0.85\linewidth]{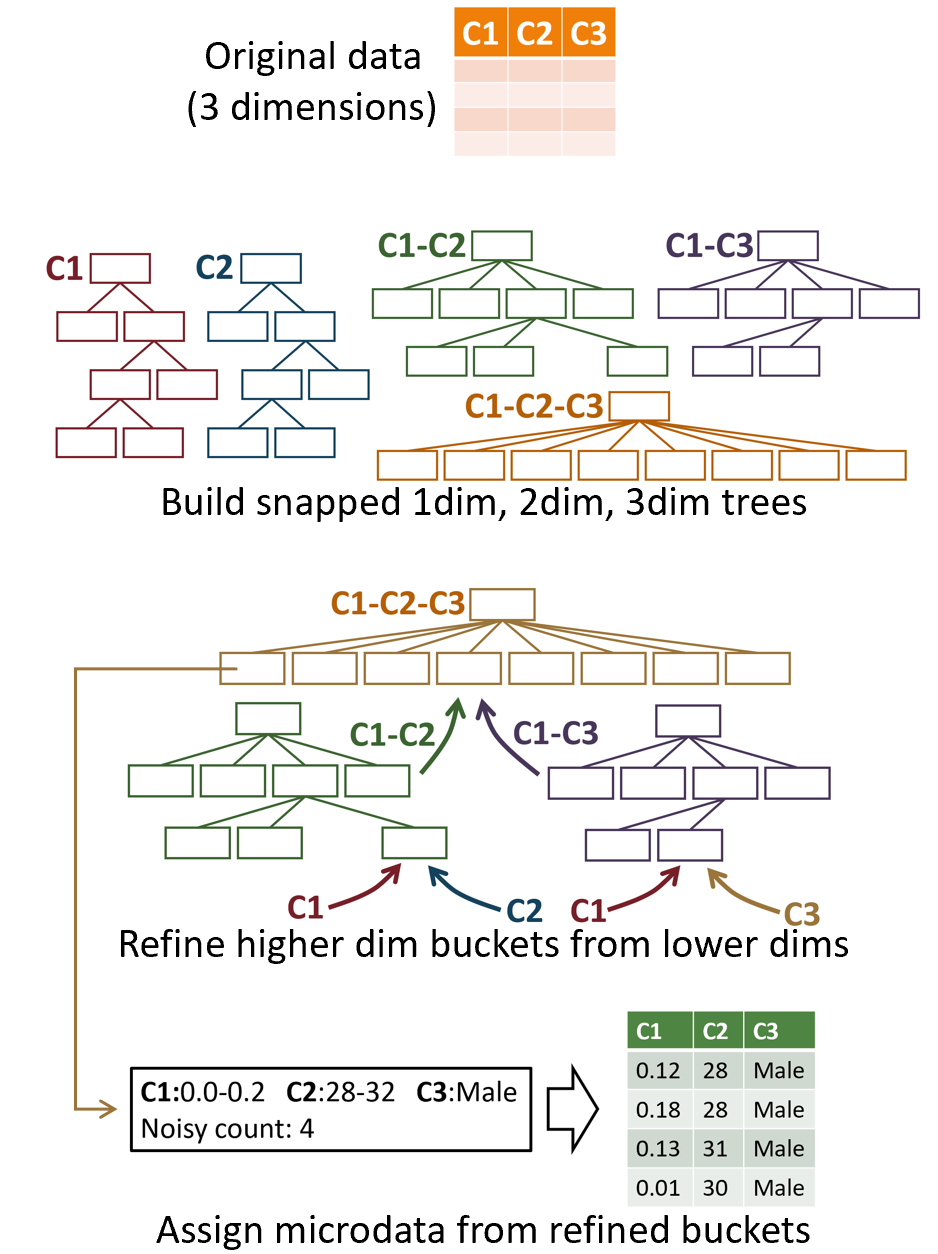}
\caption{Sketch of synthesis for low-dimension tables or sub-tables. \textnormal{}}
\label{fig:descriptive-sketch}
\end{center}
\end{figure}

\sd builds decision trees with buckets as the nodes in the trees. We build trees of all combinations of all dimensions. Lower-dimension trees have better precision (smaller ranges), but don't capture as many inter-column relationships. \sd tries to combine the precision of lower-dimensions with the inter-column relationships of higher dimensions through a process we call \textit{refining and harvesting} (\S \ref{sec:harvest}). This is illustrated in Figure~\ref{fig:descriptive-sketch}.

Of course, building trees for all combinations of columns doesn't scale well, so \sd partitions high-dimension tables into a set of lower-dimension sub-tables, synthesizing the sub-tables, and then stitching them back together (\S \ref{sec:divide}). This overall process is illustrated in Figure~\ref{fig:ml-sketch}.

\subsection{Casting to numeric}
\label{sec:casting}

Building a decision tree in general requires only that the target data has a \mytt{greater-than} function. \sd decision trees, however, are constrained by the snapping requirement. In principle, we could define snapping criteria for each data type, but as an implementation convenience, we cast non-float data values into floats before building trees, and then cast them back when we produce microdata from the buckets.

Integers are cast to floats. Booleans are assigned values 0.0 and 1.0. Datetime values are cast to elapsed seconds from some arbitrary past date that precedes any datetime in the data (our implementation defaults to \mytt{1800-01-01 00:00:00 UTC}). Natural time boundaries (second, minute, hour etc.) might be a better choice, because a user is more likely to bin the synthetic data along these boundaries. The synthetic data is more accurate if the \sd buckets align with post-synthesis bins.

Text values are alphabetically sorted and assigned consecutive integers. When casting back, \sd preserves common prefixes within a bucket. For instance, if a bucket consists of `broad', `brock', and `broil', \sd maps this back to something like `bro*123`. This hides the individual strings (which might be distinct to a PE and therefore private) while still conveying some information about the text. There is substantial room for improvement in how \sd handles text values.

Undefined values like \mytt{NULL} are assigned float values completely outside of the snapped range of the column. So for instance, if the column is \mytt{age} and the snapped range is \mytt{[0,127]}, then \mytt{NULL} would be cast to 128.0, and the resulting snapped range would be \mytt{[0,255]}. This prevents undefined values from influencing the tree structure in the ``defined'' part of the tree.

\subsection{One-dimensional trees}

\sd starts by building a tree per column. These \textit{1dim} trees are built top-down by creating a root node (we use ``node'' and ``bucket'' interchangeably) with the appropriate snapped range, and then growing the tree by inserting one row at a time. The \textit{initial} snapped range of the root node is the smallest snapped range that encompasses the \mytt{min} and \mytt{max} values of the column and the casted \mytt{NULL} values, if any. This operation obviously requires a first pass over the data.

Each node in a \onedim tree can have two children. A node with a child is a \textit{branch}, and a node without children is a \textit{leaf}. The range for each child is $1/2$ the range of the parent. A leaf node where all rows in a bucket have the same value is a \textit{singularity}.

A new row is added with the following steps:
\begin{itemize}
   \item Search the tree to find the deepest node whose ranges contain the row's values.
   \item If the reached node is a branch, then add a child leaf, add the row to the leaf, and end.
   \item Else if the reached node is a leaf, then add the row to the reached node.
   \item If the reached node is not a singularity, and \suppress\mytt{= FALSE}, make the leaf a branch, add children, and transfer the branch's rows to the appropriate children.
\end{itemize}

After all rows are added and the tree is built, many leaves will have \suppress\mytt{= TRUE}. (These get suppressed during harvesting.) If this is the case for a child of the root node, then remove the root node and make the other child the new root node, thus halving the root node range. This prevents the possibility of determining the existence of outliers from the bucket ranges of higher-dimension trees, which can be inferred from the microdata.

\subsubsection{Optimizations}

This approach to tree building naturally tries to make buckets as precise as possible (smallest ranges). The buckets may be more precise than required by the analytic use case (e.g. salary in \$5000 bins is precise enough). In these cases, it makes sense to put a cap on bucket precision, because doing so can improve precision for other columns in higher-dimension trees.

The current implementation of \sd does not have a maximum precision feature. If it were implemented, the precision would have to be a snapped range, to avoid attacks where a user makes minor adjustments in precision to try to shift individual users from one bucket to another and make inferences as a result.

Another downside of high-precision buckets, however, is that they can produce deep trees which consume more memory and increase execution time. To mitigate this, our implementation has a two-parameter default mechanism for limiting tree size. One is a simple limit on tree depth (default 15). The other is to require a minimum number of rows to make a branch, which is expressed as a fraction of the total number of rows $N$ (default $N/10000$). When the tree depth is greater than the depth limit, and the number of bucket rows is less than the row minimum, then no more branching takes place. These defaults can be modified in the API.

\subsection{Multi-dimension trees}

A tree for $n$ columns (an \ndim tree) has $n$ ranges, one per column. The ranges for the root node are taken from the corresponding \onedim tree roots. All of the ranges of a child node are half of the corresponding ranges of the parent node. As such, there are $2^n$ possible combinations of the upper and lower halves of the ranges, and therefore up to $2^n$ children. An \ndim node is a singularity only when all of its rows have the same value in all of its columns.

\begin{figure}
\begin{center}
\includegraphics[width=0.8\linewidth]{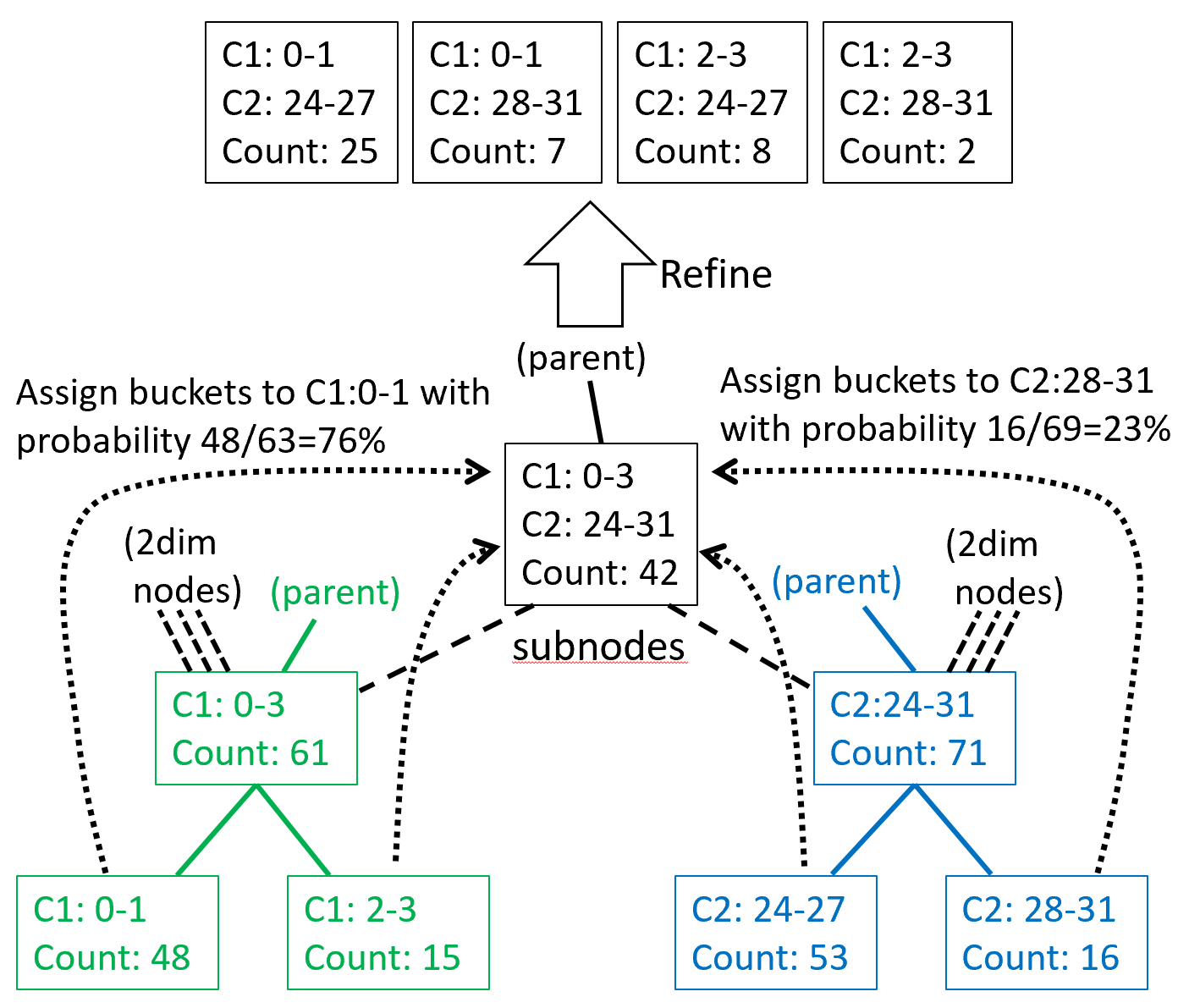}
\caption{Example of harvesting by refining. \textnormal{Higher-dimension trees capture column relationships, while lower-dimension trees have better precision. Harvesting exploits the advantages of both.}}
\label{fig:harvest}
\end{center}
\end{figure}

Each node of an \ndim tree is linked to the corresponding nodes of the lower \nonedim trees. This is illustrated in Figure~\ref{fig:harvest}, where the \mytt{2dim} node for columns C1 and C2 are linked with the \onedim nodes (called \textit{subnodes}) for the same ranges (\mytt{[0,3]} and \mytt{[24,31]} respectively).

The primary purpose of subnodes are for refining and harvesting (\S \ref{sec:harvest}). However, they are also used to prevent unnecessary branching. When a row is added to an \ndim tree, in addition to the branching criteria used for \onedim trees, branching is not done if any of the subnode row counts fall below a threshold.

We implemented different thresholds for singularity and non-singularity nodes. For non-singularity nodes, we wanted to experiment with controlling the balance between signal-to-noise and precision. A higher threshold reduces signal-to-noise at the expense of precision.

In practice we found that setting the thresholds to low values worked well. Our defaults are 15 rows for non-singularity nodes, and 5 rows for singularities.

\subsection{Harvest refined buckets}
\label{sec:harvest}

In order to generate microdata for a table with $D$ columns, we need buckets with $D$ ranges (\mytt{Ddim}). Unless D is quite small, the precision of \mytt{Ddim} buckets is poor, so we use information from \mytt{D-1dim} buckets to improve, or \textit{refine} precision. These in turn use information from \mytt{D-2dim} buckets, and so on all the way down to 1dim buckets. This overall process is called \textit{refining and harvesting} buckets.

We describe refining by way of example. The complete details can be found in the source code.

Figure~\ref{fig:harvest} illustrates refining a \mytt{2dim} leaf node and associated \mytt{1dim} subnodes. The two columns are C1 and C2. The counts are noisy counts. Since the \mytt{2dim} node has a noisy count of 42, it should produce refined buckets with counts totaling 42 rows.

The ranges of the \mytt{2dim} leaf node are \mytt{C1[0,3]} and \mytt{C2[24,31]}. The \mytt{1dim} leaf nodes are one level deeper, and so their widths are half that. Therefore, we can generate four refined \mytt{2dim} buckets from the single \mytt{2dim} leaf node.

Looking at the \mytt{1dim} tree for C1, we see that there are substantially more rows in the \mytt{[0,1]} range than in the \mytt{[2,3]} range. The \mytt{2dim} buckets should therefore have proportionally more rows in the \mytt{[0,1]} range. The distribution of rows over the C2 range is likewise skewed.

By assigning 42 rows based on the distributions of the \mytt{1dim} leaf nodes, we produce the \mytt{2dim} buckets shown at the top of Figure~\ref{fig:harvest}.  Assuming that there are one or more \mytt{3dim} trees, these four buckets would in turn be used to refine all of the corresponding \mytt{3dim} nodes.

Note that the distribution of rows among these four refined \mytt{2dim} nodes generally does not exactly match the true underlying distribution. The C1 \mytt{1dim} nodes reflects the global distribution of rows to the ranges \mytt{C1[0,1]} and \mytt{C1[2,3]}. The refine operation, however, implicitly assumes that the C1 distribution within the corresponding \mytt{C2[24,31]} range matches that global distribution. In general this will not be the case, and indeed the distribution could be quite different.

This mismatch is intentional. The principles of bucket anonymization dictates that the \mytt{2dim} buckets cannot be more precise without compromising those principles to some extent.

\subsection{Assign microdata}

Microdata is generated from the harvested \mytt{Ddim} buckets by generating the noisy count of rows for each bucket, and assigning values to the ranges of each bucket for each row. If a given range is a singularity, then the value of that singularity is assigned. If the range is not a singularity, then random values are selected from the range.

The numeric value is then cast back to its original type as described in \S~\ref{sec:casting}.

\subsection{Partition and stitch high-dimension tables}
\label{sec:divide}

When the use case is descriptive analytics, then synthesizing only a few columns normally suffices. When the use case is ML modeling of some sort, then more columns must be synthesized; at least those that are important features for the ML model if not the whole table.

\begin{figure}
\begin{center}
\includegraphics[width=0.70\linewidth]{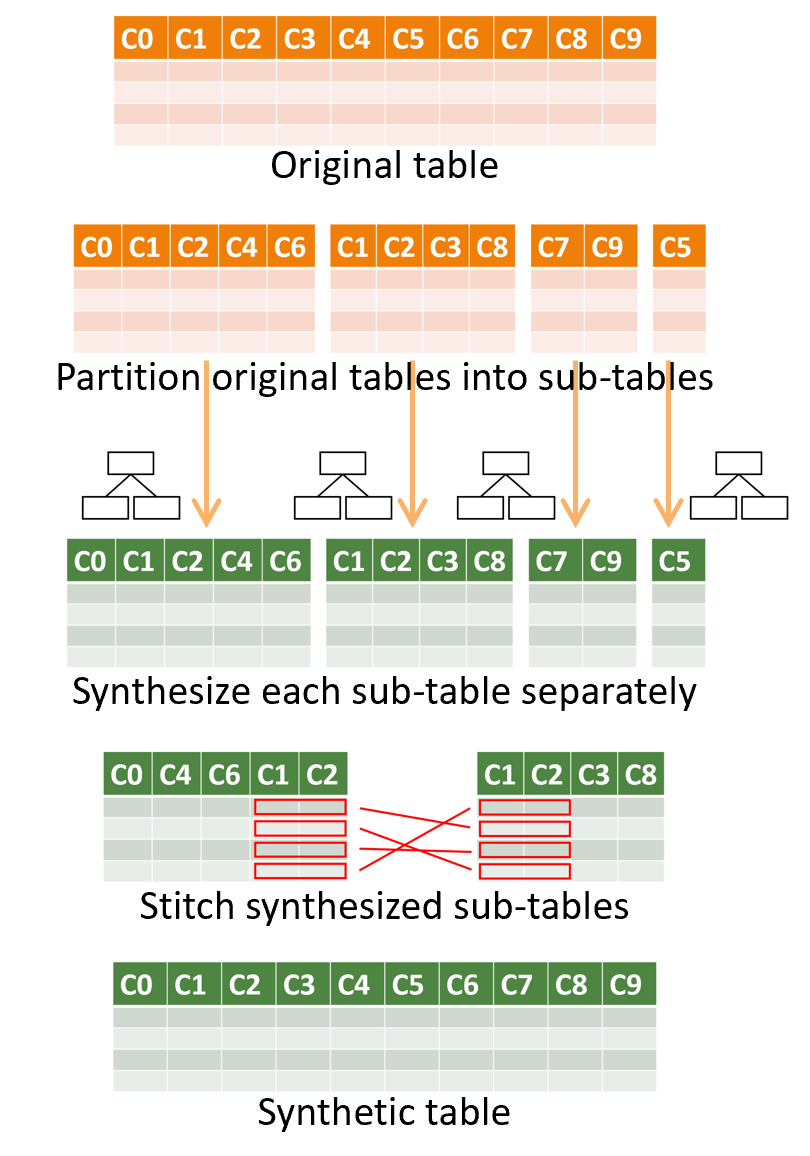}
\caption{Partitioning and stitching of high-dimension tables. \textnormal{}}
\label{fig:ml-sketch}
\end{center}
\end{figure}

Unfortunately, the approach described so far scales poorly with the number of dimensions. Therefore for high-dimension tables, it is necessary to partition the table into multiple sub-tables, synthesize each of those, and then stitch them back together. This process is illustrated in Figure~\ref{fig:ml-sketch}.

Our initial goal for \sd was to follow the one-size-fits-all \textit{generalized} approach of other synthesizers, which certainly is more convenient than generating a different synthetic table for each ML task (\textit{ML-targeted} approach). We were, however, unable to generate one-size-fits-all synthetic tables that performed as well as the best commercial product (MostlyAI), or as well as the ML-targeted approach of \sdx, which has roughly 50\% better ML efficacy than generalized \sdx.

We tried a variety of approaches to sub-table selection for generalized \sdx. Other than a brief description of sub-table selection for generalized \sd in \S \ref{sec:gen-clusters}, we focus here on ML-targeted sub-table selection.

\subsection{Sub-table selection for ML-targeted \sd}

\sd starts by doing a feature selection on the table. For this, we use \mytt{sklearn.feature\_selection.RFECV()} (recursive feature elimination with cross-validation). This produces a list of $K$ columns in order of most to least important feature.

\sd then generates one or more sub-tables from the $K$ feature columns. Each sub-table contains the ML-target column. The first sub-table contains the $C-1$ strongest features, the second sub-table contains the next $C-1$ strongest features, and so on, where $C$ is the maximum number of columns per sub-table. The default is $C=5$.

The non-feature columns are not added to any sub-tables, but rather are individually synthesized.

Note that we also experimented with variable-sized sub-tables. The idea here is that trees for columns with low cardinality are more efficient than trees for columns with high cardinality. Therefore, sub-tables with lower-cardinality columns can have more columns than sub-tables with higher-cardinality columns. In our experiments, this approach did not perform better than fixed-sized sub-tables.

\subsection{Stitching}

Each of the synthesized sub-tables has the ML-target column in common. When stitching the sub-tables together to form the complete table, this common column is used as the basis for determining which rows in each sub-table join with each other. Stitching two sub-tables produces a new sub-table with a single ML-target column. Sub-tables can be stitched in any order. Stitching two sub-tables is done as follows:

\begin{itemize}
   \item Set $N=({N_l}+{N_r})/2$, where $N$ is the number of rows for the stitched sub-table, and $N_l$ and $N_r$ are the number of rows for the two sub-tables respectively.
   \item Remove or replicate randomly selected rows from each sub-table so that they both have $N$ rows.
   \item Randomly shuffle the rows of each sub-table.
   \item Sort the rows of each sub-table on the common column.
   \item Join each row in order. The value for the common column can be taken from either sub-table.
\end{itemize}

The reason for the random shuffle is to remove any bias in row ordering that may exist after synthesis. This bias occurs whenever there are multiple rows associated with values in the ML-target column. The bias can lead to incorrect correlations between columns in different sub-tables. Shuffling removes the bias.

After the sub-tables are stitched, the rows of each remaining synthesized column (those not selected as features) are randomly removed or replicated to match the number of rows in the stitched sub-tables. The remaining columns are then randomly shuffled and stitched in.

\subsection{On the difficulty of sub-table definition for generalized \sd}
\label{sec:gen-clusters}

We tried a variety of techniques to synthesize tables that performed well for any ML model on any target column, without much success. There are a number of competing factors we need to balance when selecting sub-tables:
\begin{itemize}
   \item It is good to put high-dependence column pairs in the same sub-table. To measure dependence, we used a modified chi-square test on the buckets produced by \mytt{2dim} trees, but there may certainly be better alternatives.
   \item There is some trade-off between larger sub-tables with lower average dependence measures, and smaller sub-tables with higher average dependence measures.
   \item There is some trade-off between sub-tables with more mutual common columns used as stitch columns and fewer mutual stitch columns. More stitch columns may do a better job of retaining column relationships between sub-tables, but also serve to lower the average dependence measure within a sub-table.
   \item In any event, sub-tables should not be too large, if for no other reason than it slows down execution time. A simple way to measure ``too large'' is the number of columns. However, columns with fewer distinct values scale better than columns with more distinct values, so a better measure would take this into account.
\end{itemize}

Without getting into details, we were not able to find an evaluation function that produced good sub-tables, even assuming one exists (we tried both greedy algorithms and simulated annealing, with similar results). The ML efficacy measure given in Figure~\ref{fig:ml} (\mytt{syndiffix\_gen}) is for the simulated annealing approach, the details of which are in the Github repository.

Suffice it to say that this is still work in progress.

\section{Test setup}
\label{sec:setup}

The utility of a synthetic dataset depends very much on the use case. For the purpose of this paper, we don't try to measure utility against any particular use cases. Rather, we have two general data quality measures; data accuracy and efficacy of ML models. Note that we have not measured utility with respect to data augmentation or data bias use cases.

\subsection{Synthesis products}

We tested the following synthesis products (where version is not known, we give the month/year when the software was used):

\vspace{-2ex}
\begin{figure}[H]
\begin{small}
\begin{tabular}{llll}
    CTGAN & 0.18.0 & business source & sdv.dev/ \\
    Gretel & 7/2023 & proprietary & gretel.ai/ \\
    MostlyAI & 3.1.0 & proprietary & mostly.ai/ \\
    \sd & 1.1.0 & open source & \sdurl \\
    Synthpop & 8/2023 & open source & synthpop.org.uk/ \\
\end{tabular}
\end{small}
\end{figure}
\vspace{-3ex}

CTGAN (Conditional Tabular Generative Adversarial Network) is from the Synthetic Data Vault (SDV). SDV implemented a number of different synthetic data mechanisms, including TVAE (Tabular Variable Auto Encoder). We tested all of them, and found CTGAN to be as good as or better than the other mechanisms, and so include only CTGAN in our results. We don't know the extent to which SDV's CTGAN departs from the description in \cite{xu2019modeling}, but assume it is close.

We chose Gretel and MostlyAI because they are prominent, well-funded startups. From the website documentation, we presume that they use GAN or VAE mechanisms, possibly among other techniques, but we don't know.

Synthpop was selected because it has seen significant academic usage (\cite{nowok2016synthpop} is cited over 300 times), and because it represents a different technology base (\S \ref{sec:related}).

CTGAN, Gretel and MostlyAI were run with default settings. Where a settings choice was given between speed and accuracy, we chose accuracy. 

Synthpop was run with \mytt{method = 'cart'} (the default), \mytt{smoothing = 'spline'}, \mytt{cart.minbucket = 5}, and \mytt{maxfaclevels = 500000}. The smoothing prevented Synthpop from directly placing sampled values into the synthetic data (which would break anonymity for columns with unique values). \mytt{maxfaclevels} was set to allow Synthpop to run for columns with many distinct values. Column order matters with Synthpop. We ordered the columns from lowest to highest cardinality.

Gretel and MostlyAI were run on their company servers. The others were run locally on the same hardware (Linux machines with Intel Xeon processors and 256GB or more of memory).

Note that we weren't able to synthesize all of the tables with Synthpop because the jobs exceeded the maximum run time of our clusters (7 days) and were aborted. Probably with effort we could have made them run, but this was not a priority for us.

The \sd implementation we tested is written in F\# with a python wrapper used to do feature selection on ML-targeted syntheses and to determine column types.

\subsection{Datasets}

Our experiments were run on two sets of tables, which we refer to as \mytt{2col} (2 column) and \mytt{real}\footnote{\datasetsurl}. The \mytt{2col} tables are used to measure data quality, while the \mytt{real} tables are used to measure ML efficacy.

The \mytt{2col} tables consist of 24 artificially-generated 2-column tables, 12 with 7K rows and 12 with 28K rows. They have a variety of continuous and categorical columns with varying marginal distributions and levels of inter-column dependence. Each group of 12 has the same set of distributions: the only difference is the larger number of rows. These are used for marginal and pairs quality measures. By using 2-column tables, we mimic the usage style of selecting only those columns that we wish to analyze. Most of the tested mechanisms don't have the option of selecting columns.

Most of the \mytt{real} tables are taken from SDV\footnote{https://docs.sdv.dev/sdv/single-table-data/data-preparation/loading-data}, with a few from Kaggle (listed in Table~\ref{tab:datasets}). These come in a variety of shapes and sizes so as to broadly test ML efficacy.

\begin{table}
\begin{center}
\begin{small}
    \begin{tabular}{llll}
        \toprule
         Dataset   & Rows & Columns & Source \\
        \midrule
         adult & 32561 & 15 & SDV \\
         age-weight-sex & 3205 & 3 & Kaggle \\
         alarm & 20000 & 37 &  SDV \\
         BankChurners & 10127 & 22 & Kaggle \\
         census \dag & 299285 & 41 &  SDV \\
         census\_extended \dag & 65122 & 19 &  SDV \\
         child & 20000 & 20 &  SDV \\
         credit & 284807 & 30 &  SDV \\
         expedia\_hotel\_logs \dag & 1000 & 25 &  SDV \\
         fake\_hotel\_guests \dag & 1000 & 9 &  SDV \\
         insurance & 20000 & 27 &  SDV \\
         intrusion & 494021 & 41 &  SDV \\
         KRK\_v1 & 1000 & 9 &  SDV \\
        \bottomrule
    \end{tabular}
\end{small}
\end{center}
  \caption{Real datasets used for ML measures \textnormal{\dag Not included in Synthpop results.}}
  \label{tab:datasets}
\end{table}

\section{Data quality}
\label{sec:quality}

From the \mytt{2dim} tables, we measure the accuracy of individual columns (marginals) and column pairs. We use the SDMetrics\footnote{https://docs.sdv.dev/sdmetrics/} library for marginal and pairs quality measures as well as ML efficacy measures. SDMetrics is from the same group that developed the SDV CTGAN implementation.

For marginal column quality measures, we use SDMetrics' Kolmogorov Smirnov statistic for continuous data, and the Total Variation Distance for categorical data. For column pairs quality measures, we use SDMetrics' Correlation Similarity score for continuous data, and a Contingency Similarity for categorical data.

The range of scores for both marginals and column pairs quality ranges from 0.0 (worst quality) to 1.0 (best quality). Note that a measure of 1.0 does not mean that the synthetic data is a perfect replica of the original data, and in particular does not necessarily imply weak anonymity.

\subsection{Marginals and pairs column quality}

\begin{figure}
\begin{center}
\includegraphics[width=0.75\linewidth]{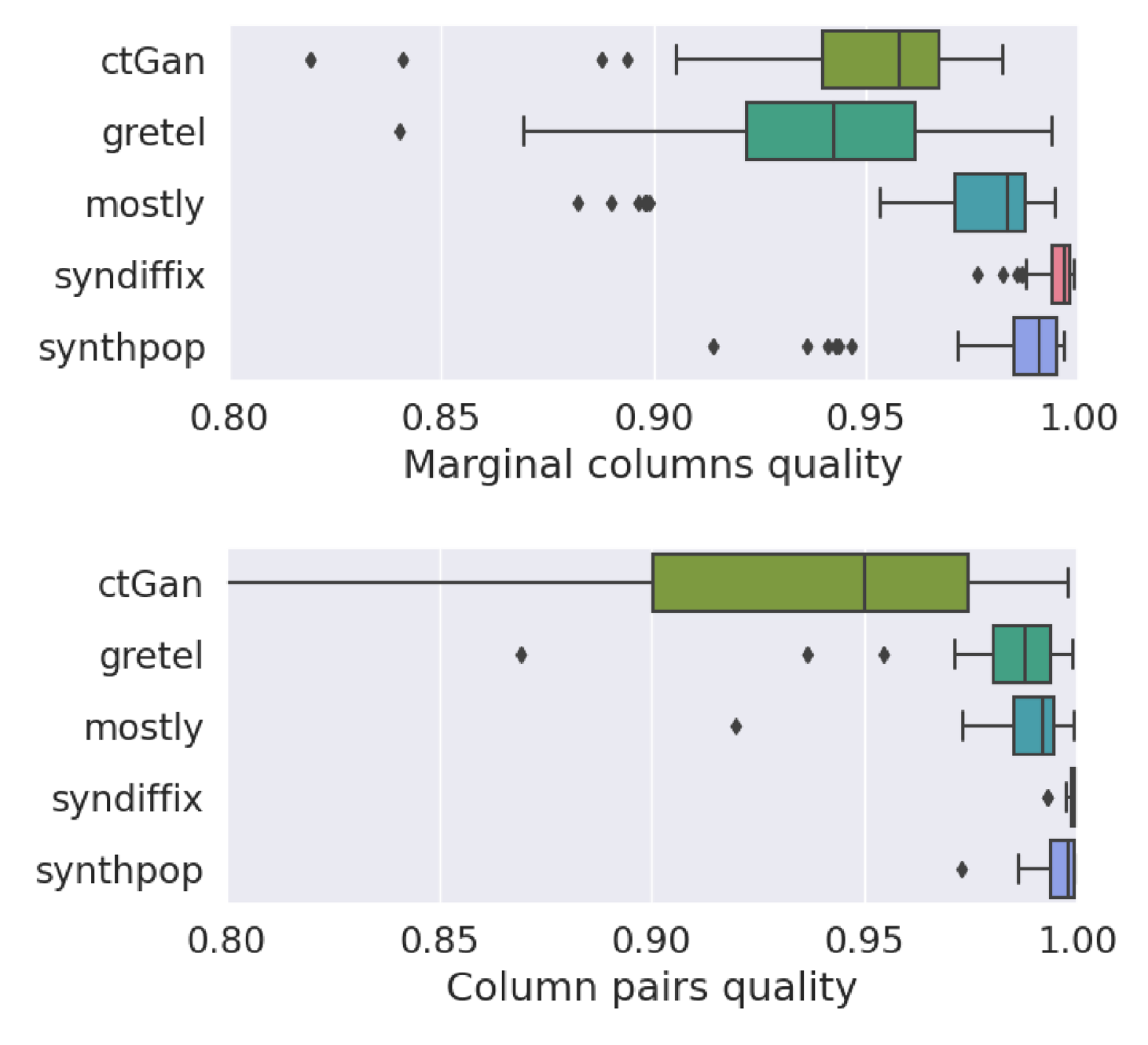}
\caption{Quality of marginals and column pairs. \textnormal{These measures are taken from 2-column tables. The marginals would be more accurate still if they were synthesized from 1-column tables.}}
\label{fig:quality}
\end{center}
\end{figure}

Figure~\ref{fig:quality} gives the marginals and pairs quality measures as boxplots over the 48 marginals and 24 pairs measures. (Note that the marginals quality measures would be better for all methods if single-column tables were used.) As can be seen, \sd has significantly better data quality.

Table~\ref{tab:quality} gives the median marginals and pairs scores, and the improvement of \sd over the other methods. Improvement is measured as $(1-{QS_{low}})/(1-{QS_{high}})$, where $QS_{low}$ is the lower quality score. This makes for instance a quality score of 0.99 two times more accurate than 0.98. Here we see that \sd is at least several factors more accurate than any of the other systems.

\begin{table}
\begin{center}
\begin{small}
    \begin{tabular}{l|llll}
        \toprule
        Method & \multicolumn{2}{c}{Marginals Quality} & \multicolumn{2}{c}{Pairs Quality} \\
            & Median & Improve & Median & Improve \\
        \midrule
        CTGAN & 0.958 & 12x & 0.950 & 79x \\
        Gretel & 0.942 & 16x & 0.988 & 19x \\
        MostlyAI & 0.983 & 5x & 0.992 & 13x \\
        \sd & 0.996 & --- & 0.9994 & --- \\
        Synthpop & 0.991 & 2.6x & 0.998 & 3x \\
        \bottomrule
    \end{tabular}
\end{small}
\end{center}
  \caption{Median quality measure for marginals and pairs quality. \textnormal{\textit{Improve} is the improvement factor of \sd over the other methods.}}
  \label{tab:quality}
\end{table}

\begin{figure}
\begin{center}
\includegraphics[width=1.0\linewidth]{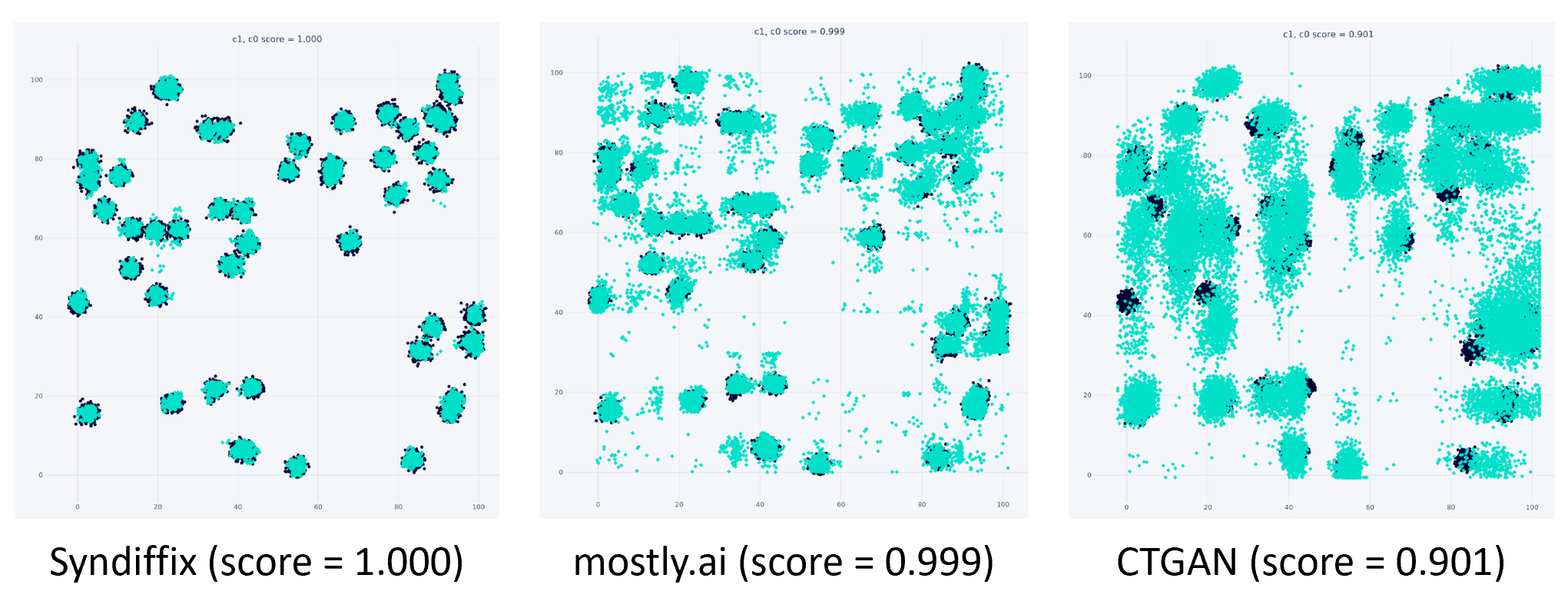}
\caption{Visualization of pairs quality measure. \textnormal{This shows that 1) small differences in the Kolmogorov Smirnov measure can make a significant difference data accuracy, and 2) a perfect Kolmogorov-Smirnov measure does not imply perfect synthesis.}}
\label{fig:scatters}
\end{center}
\end{figure}

Of course, rather than view 0.99 as being twice as accurate as 0.98, one could regard it as only being 1\% more accurate. To visualize what a small difference in this measure means, Figure~\ref{fig:scatters} plots the original and synthetic datapoints for one of the \mytt{2dim} tables for \sdx, MostlyAI, and CTGAN. Here we see that even an 0.1\% drop in the quality score leads to substantially less accuracy, at least visually.

\subsection{NYC taxi data example}

To further illustrate the practical impact of the accuracy difference, we synthesized one day (slightly over 1M rides) of the NYC taxi data using \sdx, MostlyAI, and CTGAN. We synthesized three columns; ride start hour, start latitude, and start longitude. Note that we pre-processed the time column by rounding to the nearest hour (which improves the location accuracy).

\begin{figure*}
\begin{center}
\includegraphics[width=1.0\linewidth]{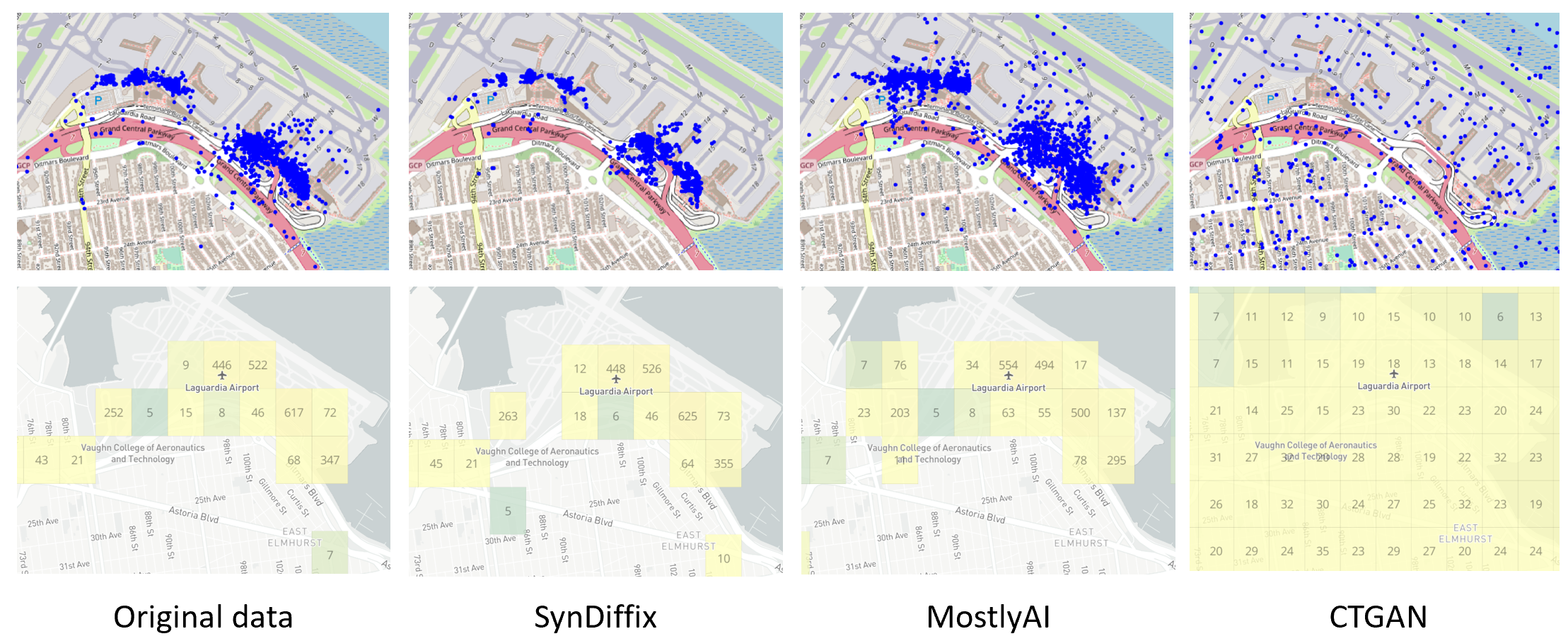}
\caption{Visualization taxi data quality. \textnormal{While visually the location of individual MostlyAI rides is pretty accurate, the heatmap data reveals that the volume of rides is much less accurate compared to \sdx.}}
\label{fig:taxi}
\end{center}
\end{figure*}

Figure~\ref{fig:taxi} zooms in on La Guardia airport, showing both a scatterplot of rides during one hour, and a heatmap generated from the synthetic data. The heatmap shows the number of rides in each location box, discarding boxes with fewer than 5 rides for visual clarity.

From the scatterplot, we see that both \sd and MostlyAI do a pretty good job of capturing location. CTGAN is very poor. MostlyAI tends to spread the locations out, while \sd tends to tighten them. 

If we look at the heatmaps, however, we see that \sd is substantially more accurate than MostlyAI. For \sdx, all of the counts are within 11 of the original data, and most of the counts are within 5. For MostlyAI, some counts are off by more than 100, and many by more than 50.

\section{ML Efficacy}
\label{sec:ml-efficacy}

To measure ML efficacy, we generated ML prediction models from the \mytt{real} tables using a variety of modeling methods over a variety of target columns. For both generating models and measuring efficacy, we used the tools from SDMetrics. SDMetrics uses the \textit{F1 test score} for binary and categorical data, and the \textit{coefficient of determination} for continuous data.

The ML models supported by SDMetrics are

       \hspace{0.5cm}\mytt{BinaryAdaBoostClassifier}

       \hspace{0.5cm}\mytt{BinaryLogisticRegression}

       \hspace{0.5cm}\mytt{BinaryMLPClassifier}

       \hspace{0.5cm}\mytt{MulticlassDecisionTreeClassifier}

       \hspace{0.5cm}\mytt{MulticlassMLPClassifier}

       \hspace{0.5cm}\mytt{LinearRegression}

       \hspace{0.5cm}\mytt{MLPRegressor}

We only want to measure ML models that perform reasonably well on the original tables. To identify these, we initially ran every appropriate ML model on every column, and selected only those that scored 0.7 or better. This resulted in 286 ML-model/target-column combinations.

Prior to synthesizing tables, we partitioned the tables into \textit{training} and \textit{test} tables, using 70\% of the data for the training tables. We then synthesize the training tables, and measure ML efficacy using the test tables.

\begin{figure}
\begin{center}
\includegraphics[width=0.75\linewidth]{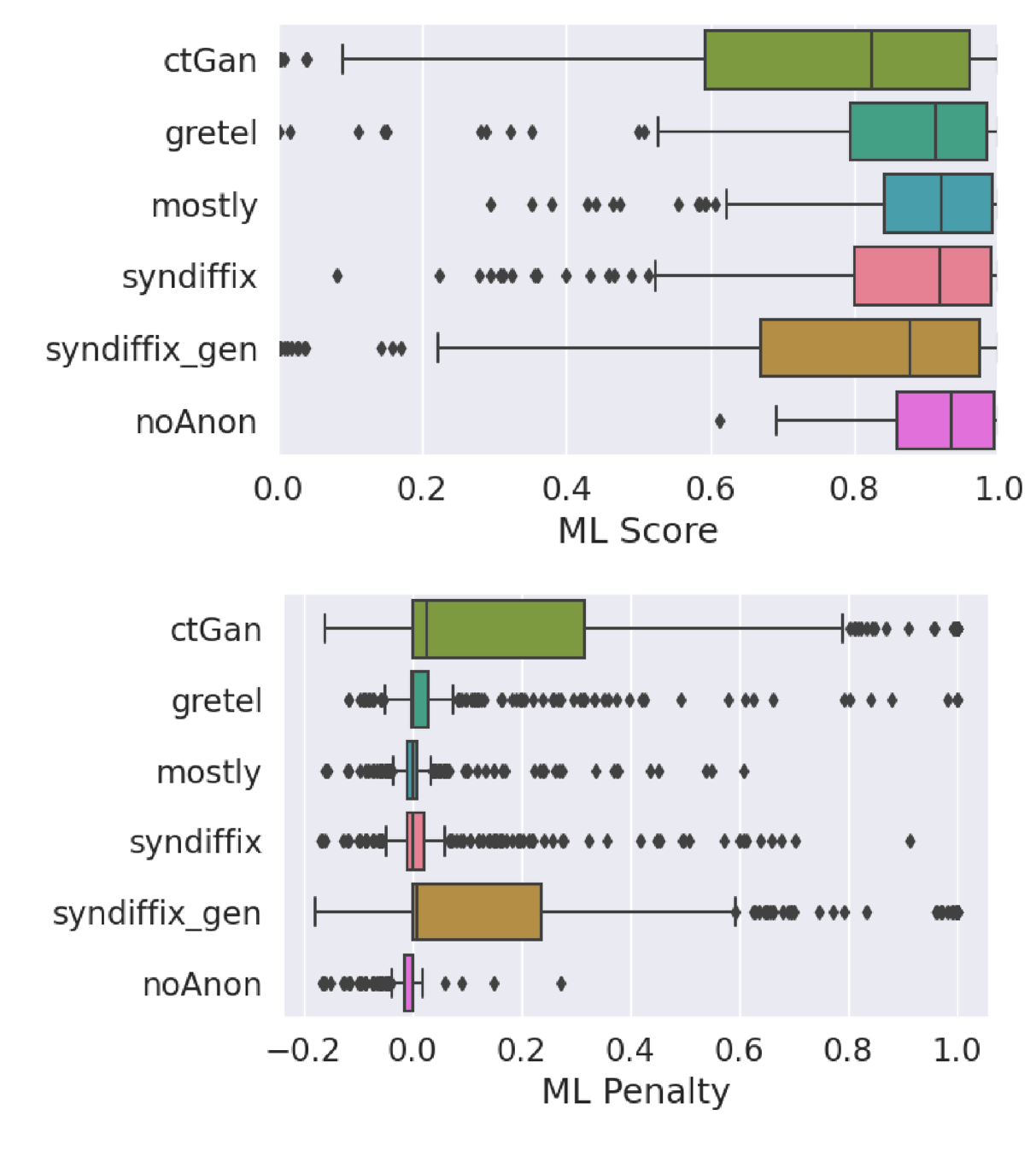}
\caption{ML efficacy. \textnormal{The ML Score is the F1 test score (categorical) and the Coefficient of Determination (continuous). ML Penalty measures the loss of efficacy from synthesis (lower is better). `syndiffix' is ML-targeted, while `syndiffix\_gen' is Generalized. `noAnon' is the original data, and is shown for calibration.}}
\label{fig:ml}
\end{center}
\end{figure}

For \sdx, we ran both the ML-targeted approach (labeled `syndiffix' in Figure~\ref{fig:ml}) and the generalized approach (labeled `syndiffix\_gen'). For the ML-targeted case, we generated a separate synthetic table for each target column. This results in 128 synthesized tables (versus 13 for the other synthesis methods).

Recall that \sd does feature selection and synthesizes sub-tables from feature columns, but synthesizes non-feature columns individually. As a result of this, we often can get better ML measures if we limit the ML model to only the feature columns (the non-feature columns only add noise to the models). We therefore took two measures for each ML model, one which includes all columns, and one which includes only feature columns, and selected the best measure. To keep our measures apples-to-apples, we did this for all of the synthesis methods.

Finally note that some of the ML models are influenced by random factors, so we ran each model 20 times and selected the best measure.

Figure~\ref{fig:ml} shows boxplots for measures of the the 286 ML models for each synthesis method. It shows both the absolute measures, and the \textit{ML Penalty}, which measures how much worse the synthesis model is compared to the corresponding model on the original data. ML Penalty is computed as $(S_{orig} - S_{syn})/max(S_{orig}, S_{syn})$, where $S_{orig}$ is the score of the model on the original data, and $S_{syn}$ is the score of the model on the synthetic data. The median ML Scores, and the improvement (or lack thereof) of \sd over the other methods, is shown in Table~\ref{tab:ml}.

\begin{table}
\begin{center}
\begin{small}
    \begin{tabular}{l|ll}
        \toprule
            & Median & Improve \\
        \midrule
        CTGAN & 0.82 & 2.1x \\
        Gretel & 0.912 & 1.07 (7\% better) \\
        MostlyAI & 0.920 & -1.04 (4\% worse) \\
        \sd (ML-targeted) & 0.917 & --- \\
        \sd (Generalized) & 0.88 & 1.48 (48\% better) \\
        Original (noAnon) & 0.935 & -1.26 (26\% worse) \\
        \bottomrule
    \end{tabular}
\end{small}
\end{center}
  \caption{Median ML Score. \textnormal{This gives the median ML Score and the improvement of \sd (ML-targeted) over the other methods. Where \sd has a lower ML Score, we show a negative improvement.}}
  \label{tab:ml}
\end{table}

In both plots, for calibration we include the measures for the original data (\mytt{noAnon}). To produce these measures, we reran the measures over the original data. For the ML Penalty, we compare the rerun measures to the initial-run measures that were used to select the good models. From the ML Penalty scores, we see that even with best-of-20 there is still some variation due to randomness.

The median ML score for \sd is comparable to both MostlyAI and Gretel. All of them produce ML models whose median is around 25\% worse than models made with the original data. While this is not horrible, it isn't that great either. It's easy to imagine that this loss of ML efficacy could make an important difference in some use cases. The ML Scores of CTGAN and Generalized \sd are significantly worse (2x and 48\% respectively).

We don't show the results for Synthpop in Figure~\ref{fig:ml} and Table~\ref{tab:ml} because Synthpop was unable to generate all of the synthetic tables. For tables that that Synthpop could produce, the median score (for 173 ML models) is 0.87. The median score for \sd is 76\% better (for the same tables).

\section{Execution time}
\label{sec:elapsed}

\begin{figure}
\begin{center}
\includegraphics[width=0.75\linewidth]{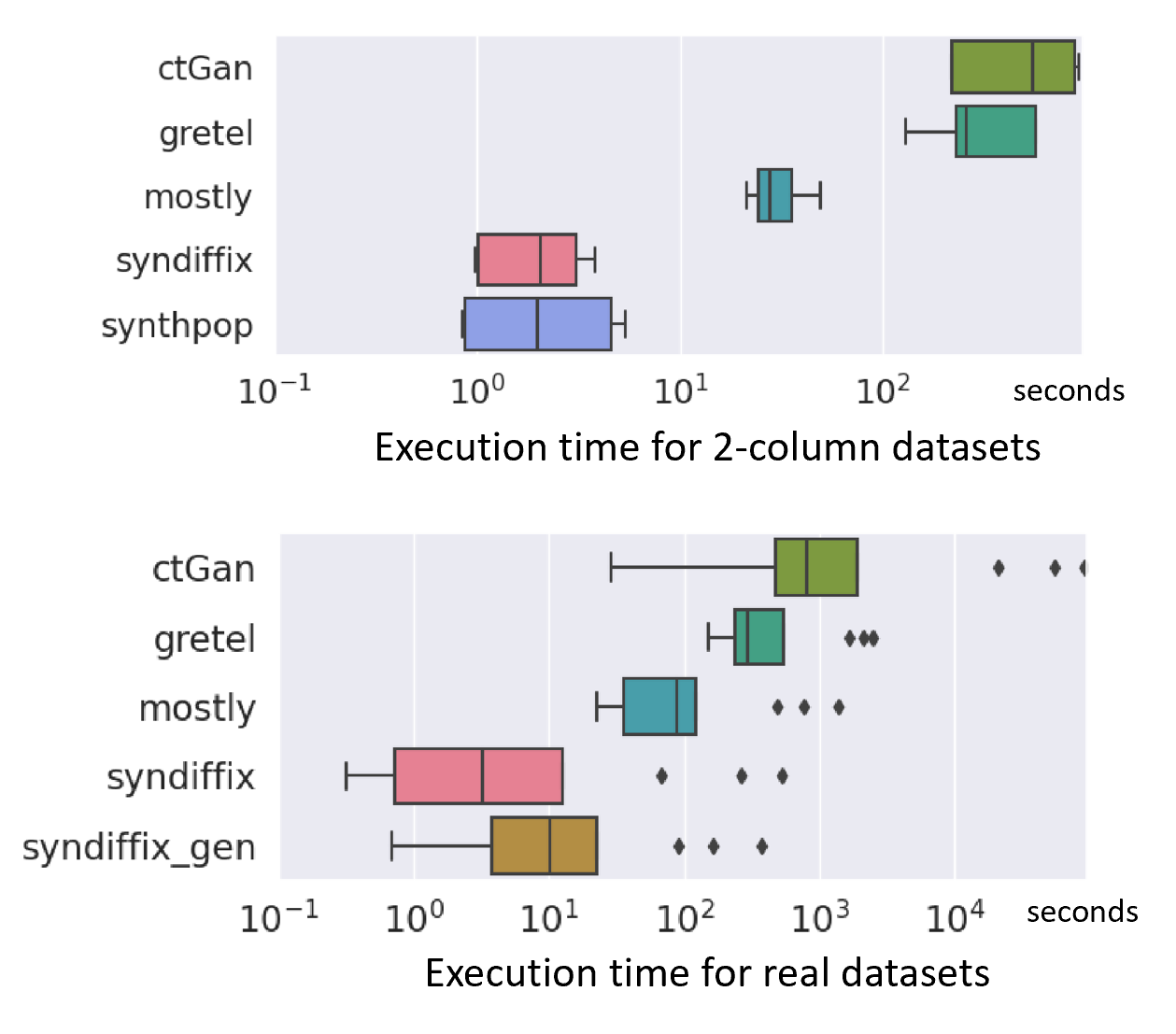}
\caption{Execution times. \textnormal{Note log scale.}}
\label{fig:elapsed}
\end{center}
\end{figure}

Figure~\ref{fig:elapsed} shows the execution times for both the \mytt{2dim} and \mytt{real} tables. Both MostlyAI and Gretel publish metadata, including execution time, alongside the synthetic data. Using this we could get measures for execution time, excluding upload and download times, from their services. Note that MostlyAI and Gretel were run on different machines than \sdx, Synthpop, and CTGAN. Note that these performance numbers are for the F\# version of \sdx, not the pure python version, which is around five times slower.

The execution times for \sd and Synthpop are similar. Synthpop execution times are only shown only for the \mytt{2dim} tables. This is because not all of the \mytt{real} tables were completed. For the tables that Synthpop did complete, the median \sd run was 7\% slower.

\sd and Synthpop are far faster than the other tools: MostlyAI is an order of magnitude slower, and Gretel and CTGAN are another order of magnitude slower still.

Note that for \mytt{real} tables, Generalized \sd is substantially slower than ML-targeted \sdx. This is primarily because Generalized \sd builds multi-dimensional tables from all the columns, whereas ML-targeted does so for only the feature columns. Note also that for \mytt{2col} datasets, we only used Generalized \sdx.

Note finally that, while our implementation is not parallelized, \sd is parallelizable. Each separate tree can be generated on a different core.

\section{Strength of Anonymity}
\label{sec:privacy}

Both \sd and Diffix are built from the same Diffix buckets: aggregates with snapped ranges, sticky noise, and suppression. Diffix has undergone substantial evaluation over multiple years, ultimately cataloging 38 different attacks and their (lack of) effectiveness. While Diffix is not exactly the same as \sdx, for the most part the vulnerability analysis for Diffix Elm \cite{francis2022diffix} applies to \sdx. We are not aware of any effective attacks against \sdx (or Diffix Elm).

The analysis in \cite{francis2022diffix} is attack based: an attack is run, normally with a goal of inferring an unknown attribute about an individual in the data. A few attacks aim to single out an individual~\cite{article29}, i.e. to predict that certain attributes of an individual are unique in the data, but primarily we are interested in inference.

The effectiveness measure in \cite{francis2022diffix} is similar to precision and recall. These measures are intuitive: precision tells an attacker how likely a prediction is to be correct, and recall tells an attacker how likely a given individual is vulnerable. It is critical, however, that precision is expressed in terms of how much \textit{better} it is from a statistical guess. An attack that achieves 50\% precision inferring sex is not effective, because a statistical guess will do just as well.

So long as \textit{precision improvement} is low \textbf{or} recall is very low, then we can regard an attack as ineffective. While it is up to policy makers to decide, we think that 50\% or lower precision improvement is safe. The lack of certainty makes the prediction of low value to an attacker and gives the target substantial deniability. We think that a recall of 1/1000 is safe because an attacker will be discouraged from trying an attack if it is so likely to fail.

The measure we use in this paper to compare the different systems is Anonymeter inference attack\footnote{https://github.com/statice/anonymeter}~\cite{giomi2022unified}. In this attack, the attacker knows some attributes of some target individual (the \textit{victim}), and wishes to learn an unknown attribute (the \textit{secret}). The attacker finds the record in the synthetic data that best matches the victim, and predicts that that secret is that of the record. This is not the most sophisticated attack (those can be found in \cite{francis2022diffix}), but it is a simple attack that anyone might think to try.

Anonymeter measures precision improvement. It has an ingenious way of approximating the statistical baseline from which improvement is measured. The dataset is randomly split into \textit{test} and \textit{control} sets. The test set is synthesized. Anonymeter randomly selects records from both the test set and the control set, and uses these as victims. By default, Anonymeter assumes that all attributes (columns) are known except one column, which is the secret column. Anonymeter uses the synthesized data to infer the secret. Anonymeter compares the precision of attacks with test set victims against attacks with control set victims.

The idea is that the precision of attacks on control set victims is the statistical baseline. Intuitively, if the secret is the attribute sex, we would expect precision on control set victims to be at least 50\%. If some of the known attributes correlate with sex, then precision on control set victims may be better. Regardless, this control precision cannot be due to privacy loss since the control victims are not from the synthesized data (the test set). It is a widely accepted principle from differential privacy that a data set cannot be responsible for privacy loss for individuals not in the data set~\cite{dwork2017exposed}.

Anonymeter measures how much more precise the test set attacks are compared to the control set attacks, and this determines precision improvement for inference\footnote{\cite{giomi2022unified} calls this ``risk'', but ``precision improvement'' is a more accurate term.}. Precision improvement is measured as $PI = (p_{test} - p_{control}) / (1 - p_{control})$, where $p_{*}$ is precision, the fraction of correct predictions to all predictions. A rough intuition is: if a value $X$ in appears in 40\% of rows for some column, and the attack achieves a precision of 70\%, then the precision improvement is 50\%.

We ran 500 attacks per table. Each inference attack selects a random column as the secret, and assumes all other columns are known. Anonymeter computes confidence bounds on its scores, and we removed all attacks where the confidence bound was greater than 0.2 as being unreliable. (These usually occurred in inference attacks on columns where a very large majority of rows have a single value.)

\begin{figure}
\begin{center}
\includegraphics[width=0.75\linewidth]{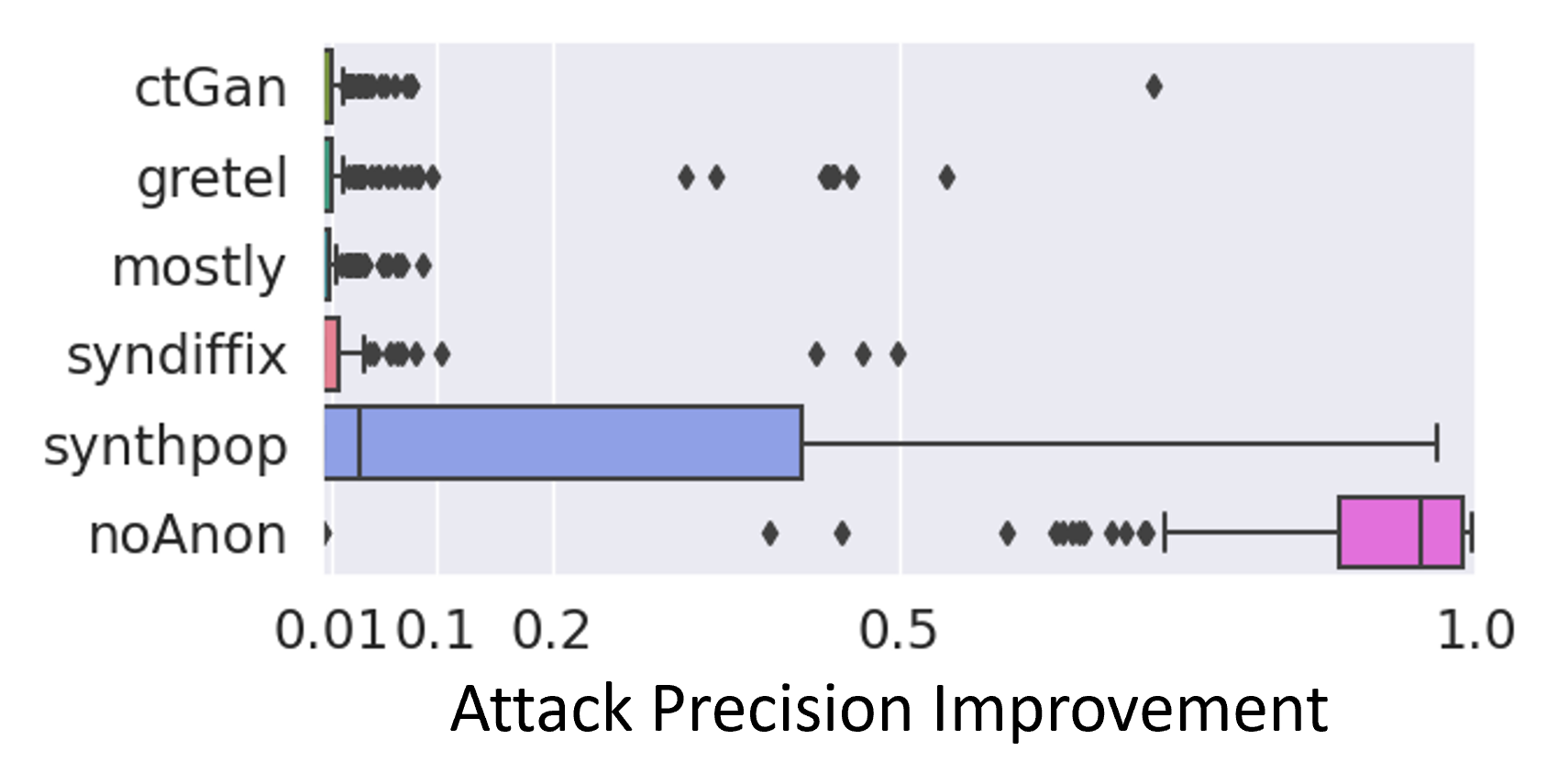}
\caption{Strength of anonymity as measured by Anonymeter. \textnormal{Attack Precision Improvement reflects how much better the attack performs compared to a statistical guess. Conservatively, any score below 0.5 can be regarded as low risk. `noAnon' is for attacks on the original data, and is shown here for calibration.}}
\label{fig:privacy}
\end{center}
\end{figure}

The results are presented in Figure~\ref{fig:privacy}. For calibration, we also present the results for attacks where the test set is not anonymized (\mytt{noAnon}). Except for Synthpop, all of the systems have very strong privacy.

One of the limitations of Anonymeter is that it doesn't try high-precision/low-recall attacks: its attacks take place on random rows and values and therefore represent 100\% recall (through sampling). As such, it doesn't find attacks that may work, for instance, on outliers. The privacy analysis of \sd in~\cite{francis2022diffix} does account for these attacks. Nevertheless, Anonymeter covers the common case of users believing they have found a match in the synthetic data and making inferences as a result.

\subsubsection{Other anonymity measures}

We don't use differential privacy as a measure of protection because a differential privacy measure does not account for the benefits of aggregation, which is an important component of \sdx. Any differential privacy measure that we might manage to produce would badly over-estimate the actual risk.

We don't use Stadler et al.~\cite{stadler2020synthetic} because its attribute inference measure does not adjust for a statistical baseline and is therefore a meaningless measure. (Besides its attribute inference measure, Stadler et al. also has a membership inference measure. That measure assumes that there is a 50\% probability that the victim of an attack is a member. This is not a realistic assumption, and so the membership inference measure is also pretty much meaningless. In short, there is nothing in Stadler et al. that supports their conclusion --- that GAN-based synthetic data is not private.)

\section{Summary and Future Work}
\label{sec:future}

This paper describes \sdx, a mechanism for generating anonymous synthetic data from structured data. It measures its performance in terms of data quality, ML efficacy, and execution speed, and measures its privacy properties using Anonymeter. It compares \sd against two well-funded proprietary commercial products (MostlyAI and Gretel), and two popular open source tools (SDV's CTGAN and Synthpop).

\sd has between 3x and 80x better marginal and column pairs data accuracy. It has comparable execution time with Synthpop, and is between one and two orders of magnitude faster than the other techniques. It has comparable ML efficacy with all but CTGAN, for which \sd is about twice as good. Similar to the ML-based approaches, it has very strong anonymity. Its strength of anonymity is substantially better than Synthpop. In short, \sd matches or exceeds the other mechanisms in every measure, and substantially exceeds every other mechanism in at least one measure.

The core element of \sd is the anonymous \textit{Diffix bucket} with its properties of sticky noise on aggregate counts, snapped ranges, and low-count suppression. We believe that this is a more sound foundation for anonymity than that of the other approaches, which is, in a nutshell, overfitting avoidance. In particular, the Diffix bucket allows a given datapoint to be part of many synthetic datasets without continuously weakening anonymity. This in turn enables a tailored rather than one size fits all approach to synthesis, which allows any given synthesis to focus only on the attributes of interest.

Because of these properties, we believe that \sd is a good basis for adding many additional data attributes. Examples include event data attributes like inter-event timing and sequences, and the attributes of text fields like text length, symbol frequencies, and text structure (e.g. credit card numbers). This is one area for future work.

Another broad area for future work is to revisit all of the design decisions to find quality improvements or optimizations (including parallelism). For instance, the current design tends to overly tighten datapoints a little. There should be ways to improve on this without compromising anonymity.

Currently \sd does not excel at ML modeling. In principle we see no particular reason that \sd can't be as accurate at building ML models as it currently is at data quality. Note also that we have not tested \sd for the use cases of data augmentation or changing data bias. There is substantial future work here.

We need to exercise \sd in real use cases. In particular, there are failure modes whereby a user may not realize that \sd has not captured a particular data attribute (e.g. inter-event timing), but assumes that it has. This could lead to completely incorrect analyses. Useful would be a mechanism that allows the user to know how good an analysis is, for instance by running it in parallel with the original data and revealing roughly how similar the two analyses are.

In short, \sd is only the first version of the `Diffix bucket' style of data synthesis, and there is ample scope for continued research on this approach.

\bibliographystyle{abbrv}
\bibliography{../../masterBib/master}

\end{document}